\documentclass[10pt, a4paper, twocolumn]{article}

\usepackage{geometry}
\usepackage[utf8]{inputenc}

\usepackage[scaled=1]{helvet}
\usepackage[helvet]{sfmath}
\everymath={\sf}

\usepackage{lipsum}
\usepackage{graphicx}
\usepackage[explicit]{titlesec}
\titleformat{\section}{\bfseries\Large}{\arabic{section}. #1}{6pt}{}
\titlespacing*{\section}
{0pt}{3pt}{6pt}
\titlespacing*{\subsection}
{0pt}{3pt}{6pt}
\titlespacing*{\subsubsection}
{0pt}{3pt}{3pt}
\providecommand{\keywords}[1]{\date{\textbf{Keywords} #1}}
\usepackage[bottom]{footmisc}

\makeatletter
\def\@maketitle{%
  \newpage
  \begin{center}%
    {\LARGE \textbf{\@title} \par}%
    \vskip 1.5em%
    {\large
      \lineskip .5em%
      \begin{tabular}[t]{c}%
        \@author
      \end{tabular}\par}%
    \vskip 1em%
  \end{center}%
  {\@date
  \par}
  \vskip 1.5em}
\makeatother

\usepackage{color}
\definecolor{light-gray}{gray}{0.5}
\usepackage{fancyhdr}
\usepackage{hyperref}
\hypersetup{colorlinks=true, urlcolor=blue, linkcolor=blue, citecolor=blue}

\title{Towards personalized computer simulation of breast cancer treatment}

\author{Alvaro K\"ohn-Luque$^{a}$\thanks{Corresponding author. Email: alvaro.kohn-luque@medisin.uio.no}, Xiaoran Lai$^{a}$ and Arnoldo Frigessi$^{a}$ \\
\emph{\small 
${}^{a}$Oslo Centre for Biostatistics and Epidemiology, University of Oslo, Norway,
}
}

\begin{document}

\maketitle
\thispagestyle{fancy}

%%%%%%%%%%%
% Article %
%%%%%%%%%%%

\section{Introduction}

Cancer pathology is unique to a given individual, and developing personalized diagnostic and treatment protocols are a primary concern. Mathematical modeling and simulation is a promising approach to personalized cancer medicine. Yet, the complexity, heterogeneity and multiscale nature of cancer present severe challenges. One of the major barriers to use mathematical models to predict the outcome of therapeutic regimens in a particular patient lies in their initialization and parameterization in order to reflect individual cancer characteristics accurately. 

Here we present a study where we used multitype measurements acquired routinely on a single breast tumor, including histopathology, magnetic resonance imaging (MRI), and molecular profiling, to personalize a multiscale hybrid cellular automaton model of breast cancer treated with chemotherapeutic and antiangiogenic agents. We model drug pharmacokinetics and pharmacodynamics at the tumor tissue level but with cellular and subcellular resolution. We simulate those spatio-temporal dynamics in 2D cross-sections of tumor portions over 12-week therapy regimes, resulting in complex and computationally intensive simulations.

For such computationally demanding systems, formal statistical inference methods to estimate individual parameters from data have not been feasible in practice to until most recently, after the emergence of machine learning techniques applied to likelihood-free inference methods. Here we use the inference advances provided by Bayesian optimization to fit our model to simulated data of individual patients. In this way, we investigate if some key parameters can be estimated from a series of measurements of cell density in the tumor tissue, as well as how often the measurements need to be taken to allow reliable predictions.

\section{Methods}
\subsection{Patients and clinical data}

To develop and validate our mathematical model we used data from HER2-negative mammary carcinomas from the NeoAva cohort \cite{silwal2017}, a randomized, phase II clinical trial that evaluated the effect of bevacizumab in combination with neoadjuvant treatment regimes. Four patients were selected to belong to both arms of the trial and to have either a complete or no response at 12 weeks of treatment. Details of baseline characteristics, treatment, response and used clinical data from all patients can be found in reference~\cite{lai2019}.

\subsection{Multiscale mathematical model}
We model the response of a cross-section of tumor tissue to a combination of chemotherapeutic and antiangiogenic drugs using a multiscale hybrid cellular automaton. We combined tissue, cellular, extracellular, and intracellular dynamics and informed them by multitype, individual patient data. Figure~1 shows an overview of the model structure and the different model formalisms used. Full details are available in reference~\cite{lai2019}.

\begin{figure}[h!]
\begin{center}
	\includegraphics[width=8cm]{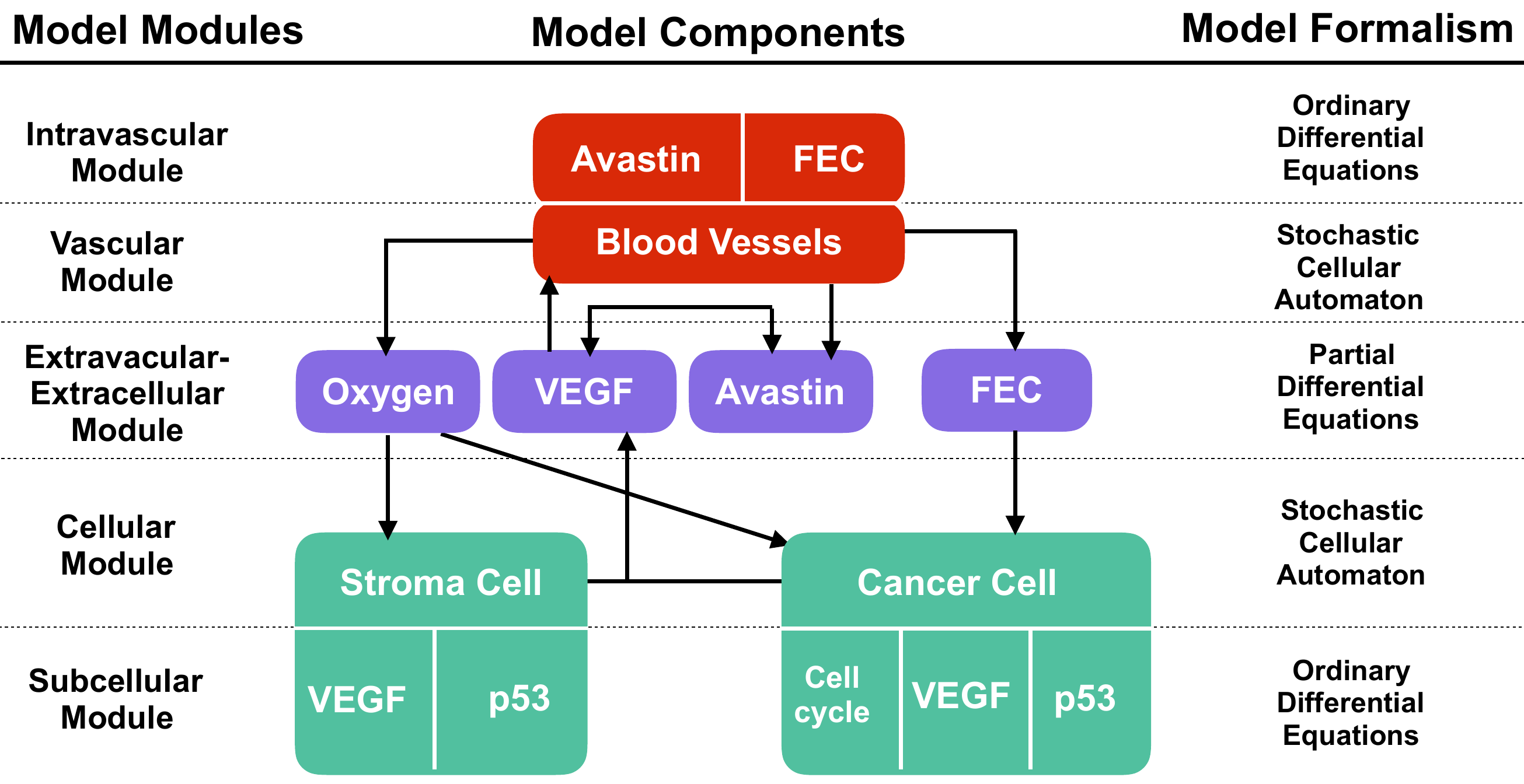}	
	\caption{Multiscale hybrid cellular automaton structure}
\end{center}
\label{fig1}
\end{figure}

\subsection{Bayesian optimization for likelihood-free inference}

For parameter inference we use an approximate Bayesian computation method called Bayesian optimization for likelihood-free inference (BOLFI) \cite{gutmann2016}. BOLFI uses Gaussian processes to model the unknown discrepancy function that compares the simulations with the data via Bayesian optimization. Once the probabilistic model has been inferred, it can be used to approximate the true posterior distribution.

\subsection{Simulation code}

The  open-source code  to  simulate  the  multiscale mathematical  model  is  written  in  Python  3.6.2  and  available  at bitbucket.org/xlai/codeavastin-py.git  

\section{Results}

\subsection{Simulation of observed treatment outcomes}
We first use our model to reproduce and elucidate the treatment outcome of four patients. For each of them, we run personalized simulations of different tumor portions under the treatment received in the trial.  Tumor screening data at baseline (histopathology, MRI and molecular data) are used to initialize and parameterize the model together with other data available in the literature. We simulate 12 weeks of the spatio-temporal dynamics of the considered tumor portion under the effect of the applied drugs. We then use MRI examinations at weeks 1 and 12 to compare with the simulated outcomes.  As an example, we illustrate in Figure 2-I a simulation of a non-responder patient with low proliferative cells able to scape the standard chemotherapy regime administered every three weeks. 

\begin{figure}[h!!]
\begin{center}
	\includegraphics[width=8.5cm,height=6.5cm]{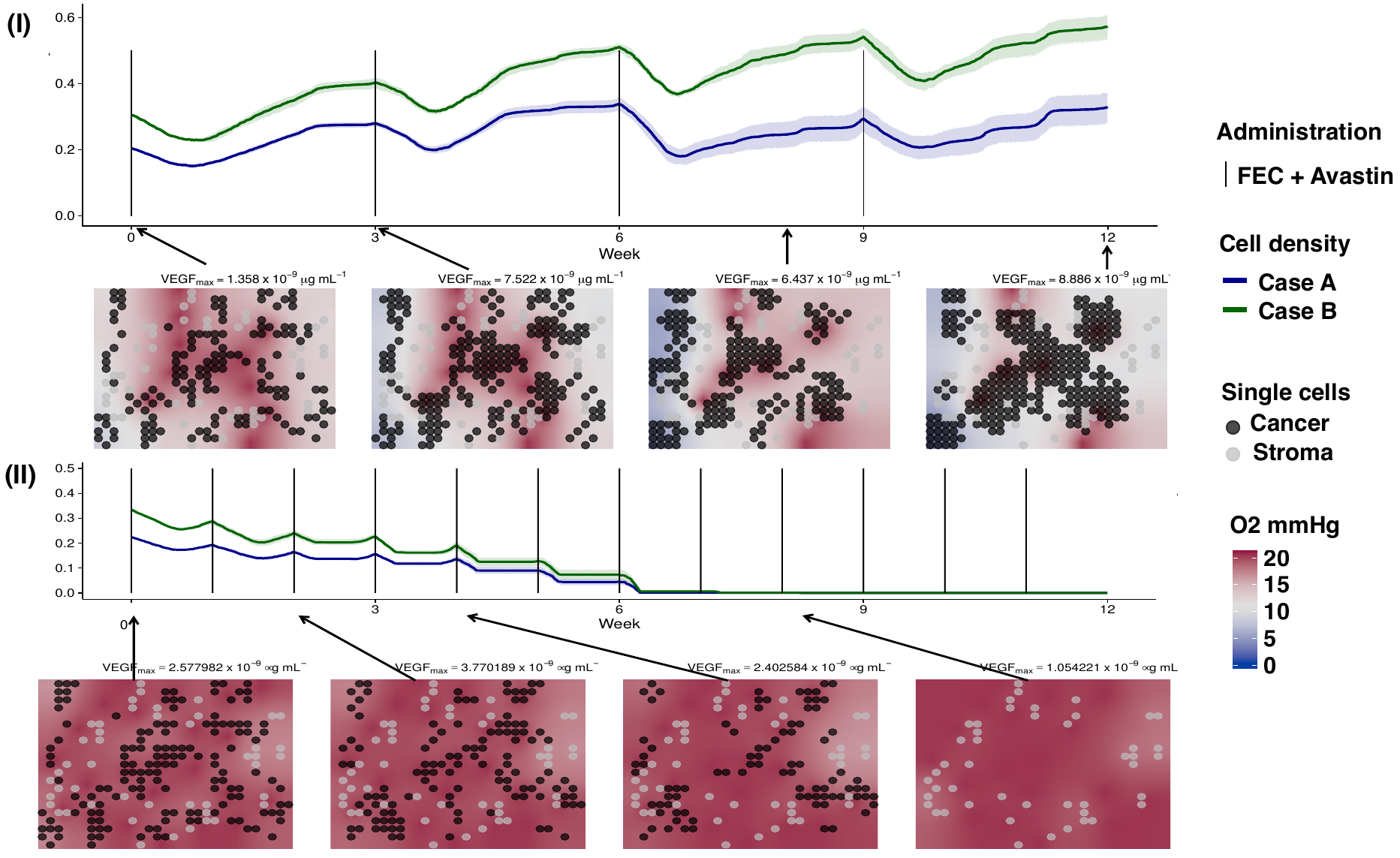}	
	\caption{Personalized treatment simulations of one patient: (I) Failed regime received in the trial. (II) Alternative regime with improved outcome. Simulated evolution of the cancer cell density starting from two observed cell configurations at week 0, case A and case B. Each line represents the average cancer cell density of 10 independent stochastic simulations with their 95\% bootstrap confidence interval.  Lower panel of each figure shows the spatial distribution of cancer and stroma cells for a representative simulation of Case B.  Background color represents oxygen pressure in mmHg.}
\end{center}
\label{fig2}
\end{figure}

\subsection{Simulation of alternative treatments}

In addition to the regimes used in the clinical study, we also tested multiple hypothetical alternative drug regimes (schedules and doses) for the non-responders. Figure 2-II shows a simulation of the most successful schedule obtained for the same non-responder patient shown in Figure 2-I. In that case the drugs doses are reduced to a third of its original ones, but administrated every week instead of every 3 weeks.  By means of the higher drug frequency,  low proliferative cells can no longer escape the chemotherapy and the tissue becomes free of cancer cells after approximately 6 weeks of therapy.
 
 \subsection{Inference and predictions}

Using the simulated number of cells from the patients as measurements, BOLFI is able to accurately infer up to three of the most relevant model parameters. In fact, parameters were estimated well enough by using only a few measurements made in the very beginning of the therapy. Importantly, this was sufficient to predict the effect of the therapy regime at the end of the 12 weeks with good accuracy.

\section{Conclusion}

The use of mathematical models and simulations to decide the best therapy for a cancer patient starts to become a possibility. For breast cancer, we have been able to develop and implement a first multiscale pharmacokinetic and pharmacodynamics model, which shows encouraging pilot runs on a few patients. Any such model needs to be informed by data from the unique patient for whom we wish to determine the best therapeutic regime. We have shown that it is possible to estimate personalized parameters with data compatible with clinical practice and use them to make reliable predictions of the effect of therapy for a unique and single patient.

%%%%%%%%%%%%%%
% References %
%%%%%%%%%%%%%%
\titlespacing*{\section}{0pt}{0pt}{-3pt}
\section{References}
\bibliographystyle{acm}
\bibliography{biblio}

\end{document}